\documentclass{ws-p9-75x6-50}
\usepackage{latexsym}
\usepackage{amsfonts}
\def\be{\begin{equation}}
\def\ee{\end{equation}}

\begin{document}

\title{Scalar-tensor $\mathbf{\sigma}$-cosmologies}

\author{Spiros Cotsakis}

\address{GEODYSYC, Department of Mathematics, University of the
Aegean,  83200, Samos, Greece\\E-mail: skot@aegean.gr}

\author{John Miritzis}

\address{Department of Marine Sciences, University of the Aegean, Mitilene 81100, Greece\\
E-mail: john@env.aegean.gr}

\maketitle

\abstracts{We show that the scalar-tensor $\sigma$-model action is
conformally equivalent to general relativity with a minimally
coupled wavemap with a particular target metric. Inflation on the
source manifold is then shown to occur in a natural way due both
to the arbitrary curvature couplings and the wavemap
self-interactions. }

\section{Scalar-tensor $\sigma$-models}

Let $(\mathcal{M}^{m},g_{\mu \nu })\;$ be a spacetime (source) manifold, $(%
\mathcal{N}^{n},h_{ab})\;$ Riemannian (target) manifold and a $\mathcal{C}%
^{\infty }$ map $\phi :\mathcal{M}\rightarrow \mathcal{N}.$ We may think of
the scalar fields $\phi ^{a},a=1,\dots ,n$, as coordinates parametrizing the
Riemannian target. Our starting point is the general action functional
\begin{equation}
S=\int_{\mathcal{M}}L_{\sigma }dv_{g},\;\;dv_{g}=\sqrt{-g}dx,
\end{equation}
where
\begin{equation}
L_{\sigma }=A(\phi )R-B(\phi )Tr_{g}(\phi ^{\ast }h)=A(\phi )R-B(\phi
)g^{\mu \nu }h_{ab}\partial _{\mu }\phi ^{a}\partial _{\nu }\phi ^{b},
\label{1b}
\end{equation}
where $A,B$ are arbitrary $\mathcal{C}^{\infty }\;$ functions of $\phi $. We
see that $S$ has arbitrary couplings to the curvature and kinetic terms and
we call it the scalar-tensor $\sigma $-model or, the scalar-tensor wavemap
action. Such a theory includes as special cases many of the scalar field
models considered in the literature (e.g.,\cite{1,2,3}). Under compact
variations of the families $g_{(s)}$ and $\phi _{(s)},\;s\in R\;$ where $%
\dot{\psi}(s)=[\partial \psi _{(s)}/\partial s]_{s=0}$, the Action
Principle, $\dot{S}=0,$ leads to the system,
\begin{eqnarray}
G_{\mu \nu } &=&\frac{B}{A}h_{ab}\left( \phi _{,\mu }^{a}\phi _{,\nu }^{b}-%
\frac{1}{2}g_{\mu \nu }g^{\rho \sigma }\phi _{,\rho }^{a}\phi
_{,\sigma }^{b}\right)  \\ \nonumber &+&\frac{1}{A}\left( \nabla
_{\mu }\nabla _{\nu }A-g_{\mu \nu }\Box _{g}A\right)  \\
\Box_{g}\phi ^{a} &+&\bar{\Gamma}_{bc}^{a}g^{\mu \nu }\partial
_{\mu }\phi ^{b}\partial _{\nu }\phi
^{c}+\frac{1}{2}RA^{a}=0,\quad \bar{\Gamma}=\Gamma (h)+C,
\end{eqnarray}
where $A_{a}=\partial A/\partial \phi ^{a},$ $C_{bc}^{a}=(1/2)\left( \delta
_{b}^{a}B_{c}+\delta _{c}^{a}B_{b}-h_{bc}B^{a}\right) $ and $B_{a}=\partial
\ln B/\partial \phi ^{a}$. Without loss of generality we may perform a
conformal transformation on the target metric, $\tilde{h}=Bh$, to find $\bar{%
\Gamma}=\Gamma (h)$ and so we set from now on $B=1$ in Eq. (\ref{1b}) and
drop the tilde on $h$. Under a conformal transformation of the source
manifold and the target metric redefinition,
\begin{equation}
\tilde{g}=A(\phi )g,\quad \pi _{ab}:=\frac{3}{2A^{2}}A_{a}A_{b}+\frac{B}{A}%
h_{ab}=:Q_{a}Q_{b}+\not{h}_{ab}  \label{conformal}
\end{equation}
(and dropping from the beginning the $\stackrel{\sim }{\Box }\ln \Omega $
term as a total divergence), the original scalar-tensor $\sigma $-model
action (1)-(2) becomes that of a wavemap minimally coupled to the Einstein
term
\begin{equation}
\tilde{S}=\int_{\mathcal{M}}\tilde{L}_{\sigma }dv_{\tilde{g}},\quad \tilde{L}%
_{\sigma }=\sqrt{\tilde{g}}\left( \tilde{R}-\tilde{g}^{\mu \nu }\pi
_{ab}\partial _{\mu }\phi ^{a}\partial _{\nu }\phi ^{b}\right) .
\end{equation}
This result shows that all couplings of the wavemap to the curvature are
equivalent. Varying this conformally related action, $\dot{\tilde{S}}=0$, we
find the Einstein-wavemap system field equations for the $\tilde{g}$ metric
and involving the $\pi _{ab}$ metric namely,
\begin{eqnarray}
\tilde{G}_{\mu \nu } &=&\pi _{ab}\left( \phi _{,\mu }^{a}\phi _{,\nu }^{b}-%
\frac{1}{2}\tilde{g}_{\mu \nu }\tilde{g}^{\rho \sigma }\phi _{,\rho
}^{a}\phi _{,\sigma }^{b}\right)  \\
\stackrel{\sim }{\Box }_{\tilde{g}}\phi ^{a} &+&D_{bc}^{a}\tilde{g}^{\mu \nu
}\partial _{\mu }\phi ^{b}\partial _{\nu }\phi ^{c}=0,\quad D=\Gamma (%
\not{h})+T,
\end{eqnarray}
with $T_{abc}=\partial _{c}Q_{ab}+\partial _{b}Q_{ac}-\partial _{a}Q_{bc}$
and $Q_{ab}=Q_{a}Q_{b}$.

\section{$\sigma$-Inflation}

Let us now assume that the source manifold $(\mathcal{M}^{m},g_{\mu \nu })\;$
is the 4-dimensional flat FRW model in the original scalar-tensor wavemap
theory (1)-(2). After the conformal transformation (\ref{conformal}), the
Friedman equation is $H^{2}=\frac{1}{3}T_{WM}^{00}$, where the 00-component
of the energy-momentum tensor of the wavemap is given by
\begin{equation}
T_{WM}^{00}=\frac{1}{2}\pi _{ab}\dot{\phi}^{a}\dot{\phi}^{b}.
\end{equation}
We see that the time derivative of this may change sign and
therefore we find that at the critical points of $T_{WM}^{00}$ the
universe inflates,
\begin{equation}
a=a_{0}\exp \left( \sqrt{\frac{1}{3}T_{WM,crit}^{00}}\,t\right) .
\end{equation}
This is the simplest example of a general procedure, which we call $\sigma $%
-inflation, in which inflation is driven both by the coupling $A(\phi )$
\emph{and} the self-interacting (target manifold is curved!) 'scalar fields'
$(\phi ^{a})$ which however have no potentials. This mechanism reduces to
the so-called hyperextended inflation mechanism\cite{1} when the target
space is the real line. On the other hand, when the curvature coupling $%
A(\phi )$ is equal to one, $T_{WM}^{00}$ can have no critical points since
it is always positive and so we have no inflationary solutions. In this case
we obtain the so-called tensor-multiscalar models\cite{4}. Inflationary
solutions become possible in this case by adding 'by hand' extra potential
terms and models of this sort abound.

Details and extensions of the present results will be given elsewhere.

\end{document}